# Stochastic optimal scheduling of demand response-enabled microgrids with renewable generations: An analytical-heuristic approach

Yang Li [a,*], Kang Li [b], Zhen Yang [c], Yang Yu [d], Runnan Xu [e], Miaosen Yang [f]

[a] School of Electrical Engineering, Northeast Electric Power University, Jilin 132012, China
[b] State Grid Qiqihar Power Supply Company, Qiqihar 161000, China
[c] State Grid Haidian Power Supply Company, Haidian 100089, China
[d] Hebei Key Laboratory of Distributed Energy Storage and Micro-grid, North China Electric Power University, Baoding 071003, China
[e] The Electric and Computer Engineering Department, Illinois Institute of Technology, Chicago 60616, US
[f] School of Chemical Engineering, Northeast Electric Power University, Jilin 132012, China

* Corresponding author. liyang@neepu.edu.cn (Y. Li)

**Abstract**

In the context of transition towards cleaner and sustainable energy production, microgrids have become an effective way for tackling environmental pollution and energy crisis issues. With the increasing penetration of renewables, how to coordinate demand response and renewable generations is a critical and challenging issue in the field of microgrid scheduling. To this end, a bi-level scheduling model is put forward for isolated microgrids with consideration of multi-stakeholders in this paper, where the lower- and upper-level models respectively aim to the minimization of user cost and microgrid operational cost under real-time electricity pricing environments. In order to solve this model, this research combines Jaya algorithm and interior point method (IPM) to develop a hybrid analysis-heuristic solution method called Jaya-IPM, where the lower- and upper- levels are respectively addressed by the IPM and the Jaya, and the scheduling scheme is obtained via iterations between the two levels. After that, the real-time prices updated by the upper-level model and the electricity plans determined by the lower-level model will be alternately iterated between the upper- and lower- levels through the real-time pricing mechanism to obtain an optimal scheduling plan. The test results show that the proposed method can coordinate the uncertainty of renewable generations with demand response strategies, thereby achieving a balance between the interests of microgrid and users; and that by leveraging demand response, the flexibility of the load side can be fully exploited to achieve peak load shaving while maintaining the balance of supply and demand. In addition, the Jaya-IPM algorithm is proven to be superior to the traditional hybrid intelligent algorithm (HIA) and the CPLEX solver in terms of optimization results and calculation efficiency. Compared with the HIA and CPLEX, the proposed method improves the MG net revenue by 10.9% and 11.9%, and reduces the user cost by 6.1% and 7.7%; our approach decreases the calculation time by about 90% and 60%.

## NOMENCLATURE

**Acronyms**

| | |
|---|---|
| MG | Microgrid |
| IMG | Isolated microgrid |
| RG | Renewable generation |
| DR | Demand response |
| CCP | Chance-constrained programming |
| MILP | Mixed integer linear programming |
| ESS | Energy storage system |
| JAYA | Jaya algorithm |
| IPM | Interior point method |
| PV | Photovoltaic |
| WT | Wind turbine |
| MT | Micro-turbine |
| EL | Equivalent load |
| PCC | Point of common coupling |
| PDF | Probability density function |
| Zn-Br | Zinc-bromine |
| SOT | Sequence operation theory |
| PSO | Particle swarm optimization |
| HIA | Hybrid intelligent algorithm |

**Symbols**

| | |
|---|---|
| $r$ | Actual light intensity |
| $r_{max}$ | Maximum light intensity |
| $\lambda_1, \lambda_2$ | Shape factor |
| $\rho$ | Integral variable |
| $P^{PV}$ | PV power outputs |
| $\xi$ | Solar irradiance |
| $\eta^{PV}$ | Conversion efficiency |
| $A^{PV}$ | Radiation area |
| $P^{PV}_{max}$ | Maximum value of PV outputs |
| $v$ | Actual wind speed |
| $k$ | Shape factor |
| $\varpi$ | Scale factor |
| $P^{WT}$ | WT power outputs |
| $v_{Rated}$ | Rated wind speed |
| $v_{in}$ | Cut-in wind speed |
| $v_{out}$ | Cut-off wind speed |
| $P^{WT}_{Rated}$ | Rated value of WT outputs |
| $P^L$ | Load active power |
| $\mu_L$ | Mean of load power |
| $\sigma_L$ | Standard deviation of load power |
| $P^{EL}$ | Power of the EL |
| $F_1$ | Microgrid cost |
| $F_2$ | User cost |
| $\omega_{rt}$ | Real-time price |
| $P^{CH}$ | ESS charge power |
| $P^{DC}$ | ESS discharge power |
| $T$ | Entire scheduling cycle |
| $t$ | Time period |
| $M_G$ | Total number of MT units |
| $\varsigma_n$ | MT starting costs |
| $\kappa_n$ | MT spinning reserve costs |
| $\zeta_n$ | MT consumption coefficients |
| $\psi_n$ | MT consumption coefficients |
| $R^{MT}_n$ | MT spinning reserve |

| | | | |
|---|---|---|---|
| $P_n^{MT}$ | MT power outputs | $F_2^{JO}$ | User cost considering both MG and user interests |
| $S$ | Start-up variables | $F_1^{IO}$ | MG net operating cost considering only MG interests |
| $U$ | State variables | | |
| $P^{LS}$ | Amount of load shedding | $F_2^{IO}$ | User cost considering only user interests |
| $SOC_t$ | ESS capacity | | |
| $\eta^{CH}$ | Charge coefficient of ESS | $\gamma$ | Preset confidence level |
| $\eta^{DC}$ | Discharge coefficient of ESS | $\partial$ | Preset load fluctuation |
| $\Delta t$ | Length of each time period | **Subscripts** | |
| $Q^{CH}$ | ESS charge reactive power | $in$ | Cut-in |
| $Q^{DC}$ | ESS discharge reactive power | $out$ | Cut-out |
| $V_t$ | Battery voltage | $min$ | Minimum value |
| $V_{max}$ | Maximum voltage of the battery | $max$ | Maximum value |
| $V_{min}$ | Minimum voltage of the battery | $Rated$ | Rated value |
| $SOC_{max}$ | Maximum capacity of ESS | $L$ | Load |
| $SOC_{min}$ | Minimum capacity of ESS | $rt$ | Real time |
| $SOC_0$ | Initial energy of ESS | $n$ | MT number |
| $SOC_{Tend}$ | Superfluous energy of ESS | $0$ | Initial value |
| $SOC_*$ | Minimum stored energy of ESS | $Tend$ | Superfluous value |
| $P_{Ress}$ | ESS reserve capacity | $*$ | Minimum stored energy value |
| $P^{CNLAOD}$ | Time-shiftable load power | $Ress$ | Reserve capacity |
| $P^{UNLAOD}$ | Non-Time-shiftable load power | $REF$ | Reference value |
| $\vartheta$ | Preset ratio of time-shiftable load | $iter$ | Current iteration |
| $P^{MOVE}$ | Amount of the movement of the time-shiftable load power | $itermax$ | Maximum iteration |
| | | **Superscripts** | |
| $q$ | Discrete step size | $PV$ | Photovoltaic |
| $\omega^{TOU}$ | Time-of-use electricity price | $WT$ | Wind turbine |
| $N_{iter}$ | Current iteration number | $MT$ | Micro-turbine |
| $N_{itermax}$ | Maximum iteration number | $EL$ | Equivalent load |
| $P_{REF}^{EL}$ | Reference power of the EL | $CH$ | Charge |
| $\omega_{REF}$ | Reference price | $DC$ | Discharge |
| $a(i_{a,t})$ | Probabilistic sequence of PV outputs | $LS$ | Load shedding |
| $b(i_{b,t})$ | Probabilistic sequence of WT outputs | $CNLOAD$ | Time-shiftable load |
| $d(i_{d,t})$ | Probabilistic sequence of load powers | $UNLOAD$ | Non-time-shiftable load |
| $e(i_{e,t})$ | Probabilistic sequence of EL powers | $MOVE$ | Amount of movement |
| $N_{a,t}$ | Sequence length of PV outputs | $TOU$ | Time-of-use |
| $N_{e,t}$ | Sequence length of EL outputs | $JO$ | Joint optimization considering DR |
| $W_{u_{e,t}}$ | Auxiliary binary variable | $IO$ | Independent optimization without considering DR |
| $F^{JO}$ | Objective function of the joint optimization | | |
| $F_1^{JO}$ | MG net operating cost considering both MG and user interests | | |

## 1. Introduction

### 1.1 Background

In the context of the low-carbon transition towards cleaner and sustainable energy production, utilizing renewable energies has been regarded as an important means for addressing existing energy and environmental issues (Rana et al., 2021), deep decarbonizing the power sector (Chen et al., 2018), and improving system operational flexibility (Li et al., 2020). As an effective way for improving the utilization efficiency of renewable energy sources and decreasing greenhouse gas emission (Aghdam et al., 2020), microgrids (MGs) have received increasing attention from academia and industry (Farrelly and Tawfik, 2020), and they can efficiently and flexibly operate in grid-connected or isolated modes (Li and Li, 2019). However, the inherent uncertainties of renewable generations make the peak load regulation of MGs difficult to handle (Hu et al., 2019), especially for an isolated MG (IMG) without the support of the main grid (Camblong et al., 2016). On the other hand, with the gradual increase in load types, strengthening demand-side management is an effective measure to relieve power tension. Aussel et al. (2020) developed a demand-side management trilevel energy pricing model for demand-side management, which provides a powerful argument for the detailed theoretical analysis and numerical experiments of the multi-level model of demand-side management pricing. As an important means of demand side management, demand response (DR) has recently received increasing attention since it can keep the balance between supply and demand through utilizing load flexibility rather than only adjusting power generations on the source side (Ju et al., 2019). Jafari et al. (2020) investigated optimal integration of renewables, diesel generators, and demand response from pollution, economy, and reliability viewpoints by using multi-objective optimization. Nowadays, DR is becoming a hot topic in the field of renewable consumption (Li et al., 2021) and smart grid (Ajoulabadi et al., 2020). For example, an optimal scheduling approach for demand response-enabled integrated energy systems with uncertain renewables has been recently proposed for renewable energy accommodation in reference (Li et al., in press-a). Accordingly, how to coordinate demand response and uncertain renewable generations for MGs is of great importance.

**1.2 Literature review**

In recent years, renewable generation (RG) uncertainties in microgrid scheduling have been taken into account in some previous works by using different methods, such as robust programming and stochastic programming. Wang et al. (2017a) proposed a multi-objective integrated scheduling method based on robust multi-objective optimization. Some previous works have used robust programming to deal with the uncertainties in multi-microgrids. Among these, Li et al. (2018a) proposed a multi-microgrids scheduling model with consideration of the cooperation of energy and reserve scheduling. Liu et al. (2017a) presented a distributed robust scheduling approach considering renewable generation uncertainties in a market environment. Qiu et al. (2018) established a two-stage robust optimization model, which not only considers the interaction between the utility and supply levels in AC/DC hybrid multi-microgrids, but also the tie-line disconnection uncertainty. It can be seen from the above literature that robust programming is a promising method to analyze several uncertain scenarios in the worst case. The solution of this method is feasible and the best definition of uncertainty set (Hosseinzadeh and Salmasi, 2015). However, the solution obtained is usually conservative because it is a hedge against the worst case (Xiang et al., 2016). At the same time, stochastic programming is another effective method to deal with the uncertainty of renewable energy resources in microgrids. Considering energy storage and user experience, a multi-objective scheduling model has been put forward for an isolated microgrid with renewables by using chance constrained programming (CCP), a branch of stochastic programming, in literature (Li et al., 2019a). In order to increase operating profit and improve the net power status of the energy-sharing network, Liu et al. (2018) established a day-ahead scheduling model based on stochastic programming, which takes into account the uncertainty of photovoltaic (PV) energy, electricity prices and load. Although stochastic programming is very effective in dealing with uncertainty, this method usually requires that the probability distribution of the control data is known or can be estimated, but in practical applications, these distributions may not be available. To address this problem, Li et al. (2019b) used discrete step transformation to obtain the cumulative distribution function of random variables and its inverse function, and thereby transformed an original CCP-based microgrid scheduling model into a readily solvable mixed integer linear programming (MILP) formulation.

As an emerging demand-side management technique, DR provides more regulatory means available for MG scheduling and it can generally be divided into incentive-based demand response and price-based demand response. There have been some pioneering works that successfully apply DR to MG scheduling. The investigations that combine demand response strategies with the uncertainty of PV outputs have been carried out. Among them, considering the intertemporal constraints and scheduling preferences of users, Mohammad and Mishra (2017) proposed a bi-level model with demand response suitable for day-ahead power market, and analyzed the potential influences of different PV and DR penetration levels on operational cost and locational marginal prices. Thomas et al. (2017) studied the optimal operation of the grid-connected intelligent building energy management system considering the uncertainty of PV outputs and the stochastic electric vehicles' driving schedule. By comparing with the method of deterministic PV outputs, it can be found that the total expected daily cost of the system using a stochastic method is lower. Wang et al. (2018) built an intelligent park microgrid model consisting of PV power generation, combined cooling heating and power systems, energy storage systems and response loads, and studied the optimal dispatch strategy of these units by considering price-based demand response. The study found that a microgrid operation optimization model that considers demand response strategies can effectively reduce operating costs and increase the utilization rate of renewable generation.

Recently, some pioneering works have attempted to combine the demand response strategy with the uncertainty of wind turbine (WT) power outputs. Li et al. (in press-a) leveraged a bi-level stochastic programming approach for scheduling of a community microgrid with an electric vehicle charging station (EVCS) and renewables. Based on the stochastic rolling-horizon optimization framework of renewable energy generation forecasts and real-time price updates in the energy market, Liu et al. (2020) studied the three types of demand response of thermostatically controlled load, deferred load and elastic load, and find that demand response can reduce the operation cost of microgrid while supporting the integration of renewable energy. Aliasghari et al. (2018) proposed a scheduling approach for a renewable microgrid with electrical vehicles with consideration of demand response and demonstrates that implementing demand response can bring profit for both microgrid and electric vehicle owners.

To summarize the unique features of the proposed method with respect to the previous works in the area, the comparison of the proposed method and related work can be seen in Table 1.

Table 1 Comparison of the proposed approach with related works

| Reference | Method | Stakeholders | | WT | PV | DR |
|---|---|---|---|---|---|---|
| | | The upper level | The lower level | | | |
| Li et al., 2019b | Chance constrained programming | Minimize the operating cost of the IMG | ---- | √ | √ | × |
| Mohammad and Mishra, 2017 | Mixed integer linear programming and optimal power flow | Minimize the compensation cost for end users | Minimize the operating cost of each trading cycle | × | √ | √ |
| Thomas et al., 2017 | Energy management system operating algorithm | Minimize the total expected daily electricity cost of the MG | ---- | × | √ | √ |
| Wang et al., 2018 | Genetic algorithm | Minimize the operating cost of the MG | ---- | × | √ | √ |
| Li et al., in press-a | Bi-level stochastic programming | Minimize the net operating cost of the MG | Minimize the EVCS net operating cost | √ | × | √ |
| Liu et al., 2020 | Stochastic rolling-horizon optimization | Minimize the operating cost of the MG | ---- | √ | × | √ |
| Aliasghari et al., 2018 | Mixed integer non-linear programming | Minimize the net cost of renewable microgrid | ---- | √ | × | √ |

| This paper | Bi-level stochastic programming | Minimize the operating cost of the IMG | Minimize the electrical cost of users | √ | √ | √ |

### 1.3 Goals

With these reported existing works, basic microgrid scheduling problems have been addressed. However, previous studies still have gaps in dealing with the uncertainty of renewable generation outputs in MGs. In addition, the reference that combines the uncertainty of RG outputs in MG with demand response strategies either only considers WT outputs or only PV outputs. To the authors' knowledge, there are very few or no studies that handle the relationship between demand response and the uncertainties of multiple renewable generations from the perspective of multi-stakeholders in the field of microgrid scheduling. In this context, the coordination of demand response and uncertainties of multiple renewables (i.e., WT and PV) are investigated in this work, and a bi-level MG scheduling model considering multi-stakeholders is proposed. It should be noted that the multi-stakeholders here refer to the MG interests and users' interests, and they seek the maximizations of their respective interests under real-time electricity pricing environments. At the same time, this paper develops a hybrid analysis-heuristic solution approach, which aims to balance the interests of IMG and users, while achieving peak load shaving. In this way, it can not only promote the consumption of renewable energy resources as much as possible, but also reduce the power generation pressure of the microgrid.

### 1.4 Contributions

The primary contributions are as follows:
1) Taking into account the inherent characteristics of multi-stakeholders and multi-objective balance, this work is the first attempt to formulate a scheduling model by leveraging the bi-level stochastic programming for coordinating demand response and uncertain renewable generations in energy management of isolated MGs, which is novel in MG scheduling methodologies. In this model, a new real-time pricing mechanism is designed to regulate the dynamic supply and demand relationship between microgrid and users.
2) To solve the proposed scheduling model, sequence operation theory (SOT) is employed for addressing the uncertainty of renewable generations and then a hybrid analytic-heuristic algorithm, called Jaya-IPM, is developed through combining the Jaya algorithm and the interior point method (IPM).
3) The presented method is successfully applied to a microgrid system and the results obtained have been compared to existing approaches. The findings and comparisons validate the performances of our approach, which provides a way of promoting the ongoing low-carbon transition towards sustainable energy production.

### 1.5 Organization

The other sections of this paper are organized as follows. The probabilistic model of intermittent renewable generations in the IMG and the probabilistic model of the load are given in Section II. After that, section III models IMG and users. Next, Section IV describes the specific solution to the problem. Section V introduces the simulation of different cases. At last, the conclusion is portrayed in Section VI.

## 2. The modeling of microgrid

Because of the renewable generation uncertainties plus the relatively small capacity of MGs, power exchanges between source and load in MGs are of the inherent nature of uncertainty, which makes the probabilistic model need to be considered in MG scheduling issues. Therefore, the probabilistic model is preferred for the presented problem statement in this work.

### 2.1 The structure of microgrid

As shown in Fig. 1, a typical MG is composed of the following components: wind turbines, photovoltaic, micro-turbines (MTs), energy storage system (ESS), and load. The point of common coupling (PCC) denotes the point of common coupling.

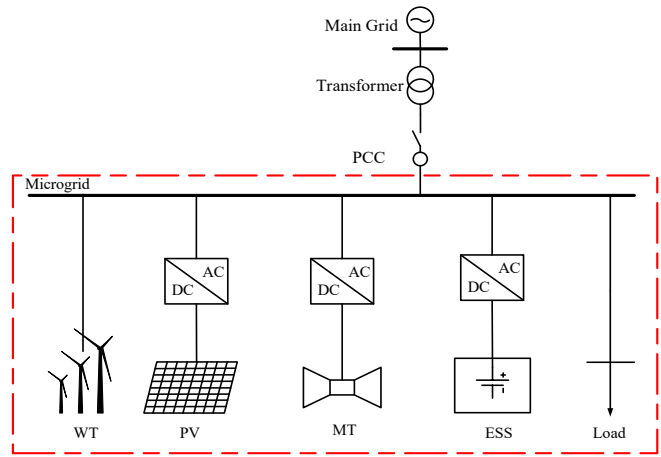

**Fig. 1.** Microgrid structure

### 2.2 Probabilistic model of photovoltaic

It's known that the solar irradiance intensity follows the Beta distribution (Li et al., 2020). In addition, the probability density function (PDF) is shown as follows (Li et al., 2019b):

$$f(r) = \frac{\Gamma(\lambda_1) + \Gamma(\lambda_2)}{\Gamma(\lambda_1)\Gamma(\lambda_2)} \left(\frac{r}{r_{max}}\right)^{\lambda_1-1} \left(1 - \frac{r}{r_{max}}\right)^{\lambda_2-1} \tag{1}$$

where $\lambda_1$ and $\lambda_2$ represent the shape factors, $r_{max}$ represents the maximum light intensity and $r$ represents the actual light intensity, $\Gamma(\cdot)$ is a gamma function $\Gamma(\lambda) = \int_0^{+\infty} \rho^{\lambda-1} e^{-\rho} d\rho$, $\rho$ represents an integral variable in the above formula (Li et al., 2018b). The relationships between the solar irradiance and the PV output $P^{PV}$ is as follows (Wang and Gooi, 2011):

$$P^{PV} = \xi \eta^{PV} A^{PV} \tag{2}$$

where $\xi$ represents the solar irradiance, $\eta^{PV}$ represents the conversion efficiency, and $A^{PV}$ is the radiation area.

Due to the linear relationships with solar irradiance, the PV output generally also obeys the Beta distribution, and its PDF is expressed as (Wang and Gooi, 2011):

$$f(P^{PV}) = \frac{\Gamma(\lambda_1) + \Gamma(\lambda_2)}{\Gamma(\lambda_1)\Gamma(\lambda_2)} \left(\frac{P^{PV}}{P^{PV}_{max}}\right)^{\lambda_1-1} \left(1 - \frac{P^{PV}}{P^{PV}_{max}}\right)^{\lambda_2-1} \tag{3}$$

where $P^{PV}_{max}$ express the maximum value of PV outputs.

### 2.3 Probabilistic model of wind turbine

Existing findings show that the PDF of wind speed can be modelled as (Xia et al., 2016):

$$f(v) = (k/\varpi)(v/\varpi)^{k-1} \exp[-(v/\varpi)^k] \tag{4}$$

where $k$ represents the shape factor (dimensionless) describing the wind speed; $\varpi$ and $v$ represent the scale factor and the actual wind speed, respectively (Li et al., 2018c).

The WT power output $P^{WT}$ and wind speed $v$ have the following relationship (Li et al., 2019b):

$$P^{WT}(v) = \begin{cases} 0 & , v < v_{in}, v > v_{out} \\ \frac{v - v_{in}}{v_{Rated} - v_{in}} P^{WT}_{Rated} & , v_{in} \leq v < v_{Rated} \\ P^{WT}_{Rated} & , v_{Rated} \leq v < v_{out} \end{cases} \tag{5}$$

where $v_{Rated}$ represents the rated wind speed, $v_{in}$ and $v_{out}$ denotes the cut-in and the cut-off wind speed respectively, and $P^{WT}_{Rated}$ is the WT rated power output.

From this, the probability density function of the $P^{WT}$ is

$$f_o(P^{WT}) = \begin{cases} (khv_{in}/\varpi P_{Rated})\left[((1+hP^{WT}/P_{Rated})v_{in})/\varpi\right]^{k-1} \times \exp\left\{-\left[((1+hP^{WT}/P_{Rated})v_{in})/\varpi\right]^k\right\}, P^{WT} \in [0, P_{Rated}] \\ 0, \quad otherwise \end{cases} \tag{6}$$

where $h = (v_{Rated}/v_{in}) - 1$.

### 2.4 Probabilistic model of loads

Load fluctuations are usually characterized by a normal distribution model, and the PDF can be described as (Xia et al., 2016):

$$f_l(P^L) = \frac{1}{\sqrt{2\pi}\sigma_L} \exp\left(-\frac{(P^L - \mu_L)^2}{2\sigma_L^2}\right) \tag{7}$$

where $P^L$ represents the load power, $\mu_L$ and $\sigma_L$ are the mean and standard deviation of $P^L$.

### 2.5 Model of equivalent loads

Since it is difficult to deal with multiple random variables at the same time, the variable named equivalent load (EL) is introduced. And thus, the EL power $P^{EL}$ is formulated as (Li et al., 2019b):

$$P^{EL} = P^L - (P^{PV} + P^{WT}) \tag{8}$$

## 3. Problem formulation

In view of the problem characteristics of multi-stakeholders and multi-objective balance, a bi-level model is put forward to coordinate demand response and renewable uncertainties. The lower- and upper-level models respectively aim to the minimization of user cost and microgrid operational cost, while real-time prices and user electricity plans are used as the links between levels.

### 3.1 The upper-level problem

#### 3.1.1 Objective function

The cost of the IMG consists of the ESS charge/discharge costs, the MT unit costs and the costs of spinning reserve and it can be formulated as

$$\min F_1 = -\sum_{t=1}^{T} P_t^{EL} \omega_{rt,t} + \sum_{t=1}^{T} (g_1(P_t^{DC}) - g_2(P_t^{CH})) + \sum_{t=1}^{T} [\sum_{n=1}^{M_G} (\varsigma_n R_{n,t}^{MT} + \kappa_n S_{n,t} + U_{n,t}(\varsigma_n + \psi_n P_{n,t}^{MT}))] \tag{9}$$

where $\omega_{rt,t}$ represents the real-time price; $g_1(P_t^{DC})$ and $g_2(P_t^{CH})$ denote the ESS discharge and charge costs during period $t$; $P_t^{CH}$ and $P_t^{DC}$ are the ESS charge and discharge powers; $T$ is a scheduling cycle; $M_G$ is the number of MT; $\psi_n$ and $\varsigma_n$ are the consumption coefficients of MT $n$ ( $n \in M_G$ ); $\kappa_n$ and $\varsigma_n$ are the spinning reserve and starting costs of MT $n$; $U_{n,t}$ and $S_{n,t}$ denote the state and start-up variables; $P_{n,t}^{MT}$ and $R_{n,t}^{MT}$ are the power output and spinning reserve of MT $n$.

### 3.1.2 Constraint conditions
#### 3.1.2.1 System power balance constraint
During scheduling, the following system power balance constraint must always be met:

$$\sum_{n=1}^{M_G} P_{n,t}^{MT} + P_t^{DC} - P_t^{CH} = E(P_t^{EL}) + P_t^{LS}, \forall t \tag{10}$$

where $P_t^{LS}$ represents the amount of load shedding; $P_t^{EL}$ and $E(P_t^{EL})$ are the EL predicted power and its expected value, respectively.

#### 3.1.2.2 Power constraint of micro-turbines
The power outputs of micro-turbines must satisfy the following constraint:

$$U_{n,t} P_{n,\min}^{MT} \leq P_{n,t}^{MT} \leq U_{n,t} P_{n,\max}^{MT}, \forall t, n \in M_G \tag{11}$$

where $P_{n,\max}^{MT}$ and $P_{n,\min}^{MT}$ are the maximum and minimum values of $P_n^{MT}$.

#### 3.1.2.3 Constraints of energy storage systems
There are many kinds of batteries, and each battery technology has its own technical and economic characteristics. Since zinc-bromine (Zn-Br) flow batteries have a relatively deeper discharge capacity, longer service life, and cheaper use costs, Zn-Br batteries have been widely used in peak-load shaving and renewable integration. Based on the above considerations, this study chooses zinc-bromine batteries to build the ESS.

**1) Equation of charge and discharge**
The evolution of the state of charge of the ESS is as follows (Chen et al., 2012):

$$SOC_{t+1} = \begin{cases} SOC_t + \eta^{CH} P_t^{CH} \Delta t \\ SOC_t - \Delta t P_t^{DC} / \eta^{DC} \end{cases}, \forall t \tag{12}$$

where $\eta^{CH}$ and $\eta^{DC}$ denote the efficiencies during the charging and discharging processes; $\Delta t$ represents the length of each time period ( $\Delta t$ =1h in this study); $SOC_t$ and $SOC_{t+1}$ represent the ESS capacity in period $t$ and $t+1$, respectively.

**2) Capacity constraint**
The state of charge of the ESS must satisfy the capacity constraint:

$$SOC_{\min} \leq SOC_t \leq SOC_{\max}, \forall t \tag{13}$$

**3) Power limits of charge and discharge**
The power limits of the ESS are formulated by

$$\begin{cases} 0 \leq P_t^{CH} \leq P_{\max}^{CH} \\ 0 \leq P_t^{DC} \leq P_{\max}^{DC} \end{cases} \forall t \tag{14}$$

$$\begin{cases} 0 \leq Q_t^{CH} \leq Q_{\max}^{CH} \\ 0 \leq Q_t^{DC} \leq Q_{\max}^{DC} \end{cases} \forall t \tag{15}$$

where $Q_t^{CH}$ and $Q_t^{DC}$ respectively denote the ESS charge and discharge reactive powers in period $t$; $P_{\max}^{CH}$ and $P_{\max}^{DC}$ are respectively the maximum values of $P_t^{CH}$ and $P_t^{DC}$; $Q_{\max}^{CH}$ and $Q_{\max}^{DC}$ are the maximum values of $Q_t^{CH}$ and $Q_t^{DC}$.

**4) Voltage constraint**
The voltage amplitude of the ESS should be limited to its allowable range:

$$V_{\min} \leq V_t \leq V_{\max}, \forall t \tag{16}$$

where $V_t$ is the voltage of the battery in period $t$; $V_{\min}$ and $V_{\max}$ denote the upper and lower limits of $V_t$.

**5) Starting and ending constraint**
For the purpose of maintaining energy balance, the initial and remaining energies of the ESS in a scheduling cycle need to be equal, which is given by

$$SOC_0 = SOC_{Tend} = SOC_* \tag{17}$$

where $SOC_*$ is the minimum stored energy of ESS, $SOC_0$ and $SOC_{Tend}$ are the ESS initial energy and its superfluous energy after a scheduling cycle.

#### 3.1.2.4 Spinning reserve constraints
Taking into account the renewable generation uncertainty, the ESS and MT units together provide the needed spinning reserves. And the spinning reserve constraints can be expressed as (Li et al., 2019b):

$$P_{n,t}^{MT} + R_{n,t}^{MT} \leq U_{n,t} P_{n,\max}^{MT}, \forall t, n \in M_G \tag{18}$$

$$P_{Ress,t} \leq \min\{\eta^{DC}(SOC_t - SOC_{\min})/\Delta t, P_{\max}^{DC} - P_t^{DC}\}, \forall t \tag{19}$$

where $P_{Ress,t}$ is the ESS reserve capacity.

Eq. (10) illustrates that $E(P_t^{EL})$ reflects the uncertainty between load and RGs. The spinning reserve furnished by the MT

units and the ESS can compensate for the discrepancy between $P_t^{EL}$ and $E(P_t^{EL})$ to achieve the balance of the electric energy.

Under some extreme conditions such as no wind at night, the joint output of RGs might be zero. In such cases, a large enough spinning reserve must be provided, which will inevitably result in expensive spinning spare costs. Of course, the probability of the presence of these conditions is very low. In this respect, the spinning reserve constraint is modeled in the form of chance constraints related to constraint-violation risks for balancing reliability and economy of MG operation (Li et al., 2019b), which is given by

$$P_{rob}\left\{\sum_{n=1}^{M_G} R_{n,t}^{MT} + P_{Ress,t} \geq (P_t^L - P_t^{WT} - P_t^{PV}) - E(P_t^{EL})\right\} \geq \gamma, \forall t \qquad (20)$$

where $\gamma$ is a confidence level.

### 3.2 The lower-level problem

Because the time-shiftable load can easily achieve peak load shaving and promote demand response, it has been extensively studied (Mohsenian-Rad H., 2015). In this work, user loads are thereby divided into two categories: time-shiftable loads and non-shiftable loads. Otherwise, the lower level is modeled to minimize the electricity cost of users. Specifically speaking, a new electricity plan is generated by superimposing the powers of both the time-shiftable load and the equivalent load; and then, the electricity plan is fed back to MG for updating its electricity price.

#### 3.2.1 Objective function

The objective function of the lower-level model is to minimize the electricity cost of users, which is given as

$$\min F_2 = \omega_{rt,t} \sum_{t=1}^{T}(P_t^{UNLOAD} + P_t^{CNLOAD}) \qquad (21)$$

where $P_t^{UNLOAD}$ and $P_t^{CNLOAD}$ denote the powers of the non-shiftable load and the time-shiftable load in period $t$, respectively; $\omega_{rt,t}$ represents the real-time price.

#### 3.2.2 Constraint conditions
##### 3.2.2.1 System power balance constraint

To guarantee that the total amount of equivalent load remains unchanged before and after the introduction of demand response, the time-shiftable loads need to satisfy the following constraints:

$$P_t^{CNLOAD} + P_t^{UNLOAD} = P_t^{EL}, \quad \forall t \qquad (22)$$

$$P_t^{CNLOAD} = \vartheta P_t^{EL}, \quad \forall t \qquad (23)$$

where $\vartheta$ ($0 < \vartheta < 1$) denotes the ratio of time-shiftable load to the total equivalent load.

##### 3.2.2.2 Upper and lower limits

The power outputs of the time-shiftable load $P_t^{CNLOAD}$ must obey the following constraint:

$$P_{t,\min}^{CNLOAD} \leq P_t^{CNLOAD} \leq P_{t,\max}^{CNLOAD}, \quad \forall t \qquad (24)$$

### 3.3 Real-time pricing mechanism

This paper proposes a real-time pricing mechanism that can automatically adjust the user electricity. The main procedures of the proposed pricing mechanism are as follows:

(1) The EL power is calculated according to Eq. (8).
(2) At the first iteration, the initial value of the real-time electricity price is set to the time-of-use electricity price.
(3) Based on the real-time price, calculate the amount of movement of the time-shiftable load $P_t^{MOVE}$ by solving the lower-level model; and thereby, the user electricity plan obtained by superimposing $P_t^{MOVE}$ and $P_t^{EL}$ is sent to the upper level (Li et al., 2018b).
(4) The price in real-time environments is determined by

$$\omega_{rt,t} = \begin{cases} \omega_t^{TOU} &, N_{iter} = 1 \\ \dfrac{P_t^{EL} + P_t^{MOVE}}{P_{REF}^{EL}} \times \omega_{REF}, & 1 < N_{iter} \leq N_{iter\max} \end{cases} \qquad (25)$$

where $N_{iter}$ and $N_{iter\max}$ respectively are the current iteration and its maximum number, $\omega_t^{TOU}$ represents the time-of-use electricity price, $P_{REF}^{EL}$ is the reference power of the EL, $\omega_{REF}$ and $\omega_{rt,t}$ are the reference and real-time prices.

(5) Repeat step (3) ~ (4) until the iterative process is terminated.

### 3.4 Compact formulation of the bi-level problem

The compact formulation of the proposed bi-level programming problem can be written as

$$\begin{cases} \min F_1 \ in \ \text{Eq.}(9) \\ s.t. \ \text{Eqs.} (10) \sim (20) \\ \min F_2 \ in \ \text{Eq.}(21) \\ s.t. \ \text{Eqs.} (22) \sim (24) \end{cases} \qquad (26)$$

## 4. Proposed solution methodology

Inspired by the solution method for bi-level programming presented in literature (Zeng et al., 2017), a new analytic-heuristic hybrid algorithm, called Jaya-IPM, is developed by combining the Jaya and the IPM considering the non-convexity of the proposed bi-level optimization model. In this algorithm, the lower- and upper- levels are addressed by the IPM and the Jaya, and scheduling schemes are determined through alternate iterations between levels. The reason for using this hybrid analytic-heuristic algorithm is based on a comprehensive consideration of solution speed and accuracy, since this combination is capable of utilizing the merits of the complementary characteristics between the analytic and heuristic algorithms when solving complex real-world optimization issues (Zeng et al., 2017). (1) On the one hand, no known effective solution method exists to solve this bi-level issue quickly and accurately although solutions are verified when given (Lachhwani and Dwivedi, 2018). (2) On the other hand, since MG scheduling is an engineering optimization problem, an approximate solution is sufficient if it can be obtained in a computationally efficient fashion for practical applications (Shrouf et al., 2014); while at the same time exact solutions are very computationally expensive when solving our proposed model, which will probably lead to failure to meet the speed requirements in practical MG scheduling applications.

### 4.1 Serialization modeling

#### 4.1.1 Basic principle of sequence operation theory

As an effective way for addressing complex uncertainties, the SOT based on sequence convolution has been successfully applied in various engineering fields. The key idea of this theory is to generate a joint probability sequence of random variables through sequence operations between discrete probability sequences, so as to solve the multiple uncertainties of random variables (Wang et al., 2017b). The advantage of this method is that the sequence operation can directly transform the chance constraint into a deterministic equivalence class without the need to solve the inverse function of the cumulative distribution function (Li et al., in press-a). More details about the SOT can be found in the related literature (Li et al., in press-b).

The probability series and their mutual operations are shown as follows.

**Definition 1.** The condition that discrete sequence $a(i)$ is a probabilistic sequence is:

$$\sum_{i=0}^{N_a} a(i) = 1, \ a(i) \geq 0, \ i = 0,1,2,...,N_a \tag{27}$$

**Definition 2.** Assuming that the length of the probabilistic sequence $a(i)$ is $N_a$, its expected value is:

$$E(a) = \sum_{i=0}^{N_a}[i \ a(i)] = \sum_{i=1}^{N_a}[i \ a(i)] \tag{28}$$

#### 4.1.2 Serialization description of renewable power outputs

Based on the SOT, $P_t^{PV}$, $P_t^{WT}$, and $P_t^L$ can be represented by $a(i_{a,t})$, $b(i_{b,t})$, and $c(i_{c,t})$ after the discretization of continuous probability distributions. Taking PV as an instance, the sequence length of the PV power output $N_{a,t}$ is given by (Li et al., 2018b):

$$N_{a,t} = [P_{\max,t}^{PV} / q] \tag{29}$$

where $P_{\max,t}^{PV}$ represents the maximum power output of the PV, and $q$ is the discretization step size.

The probabilistic sequences of the PV power outputs are shown in Table 2.

**Table 2** PV output and its probabilistic probability

| Power (kW)  | 0    | q    | … | $u_a q$   | … | $N_{a,t} q$   |
|-------------|------|------|---|-----------|---|---------------|
| Probability | a(0) | a(1) | … | $a(u_a)$  | … | $a(N_{a,t})$  |

And then, the probabilistic sequence of the PV power output is calculated by:

$$a(i_{a,t}) = \begin{cases} \int_0^{q/2} f_P(P^{PV}) dP^{PV}, & i_{a,t} = 0 \\ \int_{i_{a,t}q-q/2}^{i_{a,t}q+q/2} f_P(P^{PV}) dP^{PV}, & i_{a,t} > 0, i_{a,t} \neq N_{a,t} \\ \int_{i_{a,t}q-q/2}^{i_{a,t}q} f_P(P^{PV}) dP^{PV}, & i_{a,t} = N_{a,t} \end{cases} \tag{30}$$

The specific calculation process of this sequence is as follows: first, according to a preset discrete step size, the PDF of the PV outputs is integrated in sections; next, these integral values are used as the probabilities corresponding to different intervals; and finally, the probabilities of all intervals constitute the probability sequence of the PV outputs. In this way, a discrete sequence of PV outputs is obtained.

Similarly, the probabilistic sequence of the WT outputs and load power can also be obtained. And thereby, the expected values of the PV outputs, the WT outputs and the load powers can be calculated by

$$E(P_t^{PV}) = \sum_{i_{a,t}=0}^{N_{a,t}} i_{a,t} q a(i_{a,t}) \tag{31}$$

$$E(P_t^{WT}) = \sum_{i_{b,t}=0}^{N_{b,t}} i_{b,t} q b(i_{b,t}) \tag{32}$$

$$E(P_t^L) = \sum_{i_{d,t}=0}^{N_{d,t}} i_{d,t} q d(i_{d,t}) \tag{33}$$

### 4.2 Handling of chance constraints

#### 4.2.1 Probabilistic sequences of $P_t^{EL}$

Assuming the PV output, WT output and load power are independent in this work, the probabilistic sequences of $P_t^{PV}$ and $P_t^{WT}$ are represented as $a(i_{a,t})$ and $b(i_{b,t})$, respectively. Through addition-type-convolution operation, the probabilistic sequence of the PV and WT joint outputs $c(i_{c,t})$ is determined by (Li et al., 2020):

$$c(i_{c,t}) = a(i_{a,t}) \oplus b(i_{b,t}) = \sum_{i_{a,t}+i_{b,t}=i_{c,t}} a(i_{a,t})b(i_{b,t}), \; i_{c,t}=0,1,...,N_{a,t}+N_{b,t} \tag{34}$$

Assuming that the probabilistic sequence of load power is $d(i_{d,t})$ with the length $N_{d,t}$, through the subtraction-type-convolution operation in the SOT, the probabilistic sequence of EL power $e(i_{e,t})$ is

$$e(i_{e,t}) = d(i_{d,t}) \ominus \begin{cases} \sum_{i_{d,t}-i_{c,t}=i_{e,t}} d(i_{d,t})c(i_{c,t}), & 1 \le i_{e,t} \le N_{e,t} \\ \sum_{i_{d,t} \le i_{c,t}} d(i_{d,t})c(i_{c,t}), & i_{e,t}=0 \end{cases} \tag{35}$$

Therefore, the expected value of the equivalent load can be expressed as follows:

$$E(P_t^{EL}) = \sum_{i_{d,t}=0}^{N_{d,t}} i_{d,t} q d(i_{d,t}) - [\sum_{i_{a,t}=0}^{N_{a,t}} i_{a,t} q a(i_{a,t}) + \sum_{i_{b,t}=0}^{N_{b,t}} i_{b,t} q b(i_{b,t})] \tag{36}$$

The correspondence between the lengths $N_{e,t}$ and $e(i_{e,t})$ is shown in Table 3.

**Table 3** Probabilistic sequences of $P_t^{EL}$

| Power (kW) | 0 | $q$ | ... | $u_e q$ | ... | $N_{e,t} q$ |
|---|---|---|---|---|---|---|
| Probability | $e(0)$ | $e(1)$ | ... | $e(u_e)$ | ... | $e(N_{e,t})$ |

It can be seen from Table 3 that an EL power $u_e q$ always has a corresponding probability $e(u_e)$. The probabilistic sequence $e(i_{e,t})$ is made up of these probabilities.

#### 4.2.2 Deterministic form of chance constraint

To obtain the deterministic equivalence class of a chance constraint, a new auxiliary binary variable $W_{u_{e,t}}$ is introduced as follows (Li et al., 2019b):

$$W_{u_{e,t}} = \begin{cases} 1, & \sum_{n=1}^{M_G} R_{n,t}^{MT} + P_{Ress,t} \ge u_{e,t} q - E(P_t^{EL}) \\ 0, & otherwise \end{cases} \quad \forall t, u_{e,t}=0,1,...,N_{e,t} \tag{37}$$

From Table 3, the load power corresponding to probability $u_{e,t} q$ is $e(u_{e,t})$. Therefore, formula (20) is equivalent to

$$\sum_{u_{e,t}=0}^{N_{e,t}} W_{u_{e,t}} e(u_{e,t}) \ge \gamma \tag{38}$$

Eq. (38) suggests that for all possible EL powers, the spinning reserve capacity of the microgrid satisfies the condition that the desired confidence level is not less than the preset confidence level $\gamma$. Therefore, it can be seen that Eq. (38) and Eq. (20) are equivalent. By this means, the spinning reserve constraint in the original model has been successfully transformed from a probabilistic form into its deterministic equivalent by using the CCP.

### 4.3 Jaya algorithm

As an emerging heuristic algorithm, the Jaya has excellent stability of the solution since no algorithm-specific control parameters are required. Furthermore, the algorithm has been applied to solve various issues in the electrical engineering field, such as the configuration of renewable systems (Kaabeche and Bakelli, 2019), and feature selection for power system transient stability assessment (Li and Yang, 2017).

#### 4.3.1 Principle of the algorithm

The key point of the Jaya is that a solution ought to move towards the best solution and away from the worst one during the optimization process (Huang et al., 2017). If the $j$ variable of candidate $k$ is $X_{j,k,i}$ at the $i$th iteration, its updated value is calculated by

$$X'_{j,k,i} = X_{j,k,i} + r_{1,j,i}(X_{j,best,i} - |X_{j,k,i}|) \times r_{2,j,i}(X_{j,worst,i} - |X_{j,k,i}|) \tag{39}$$

where $X_{j,best,i}$ (/ $X_{j,worst,i}$) denotes the best (/worst) value of $j$ at the $i$th iteration; $r_{1,j,i}$ and $r_{2,j,i}$ are random numbers for $j$ during iteration $i$. If $X'_{j,k,i}$ generates a better solution, it is accepted.

#### 4.3.2 Coding scheme

This work adopts a real-integer coding scheme for facilitating optimization since the features of optimization variables (Li et al., 2018d). The continuous real variables comprise $P_n^{MT}$, $R_n^{MT}$, $P_{Ress}$, $P^{EL}$, $\omega_{rt}$, $P^{CH}$, $P^{DC}$; the discrete variables include $U_n$ and $S_n$. In this way, the initial population is randomly generated in the entire feasible domain.

### 4.4 Interior point method

Taking into account the linear programming structure of the lower-level sub-problem, the well-known primal-dual IPM is a powerful algorithm.

Step 1: Choose an initial feasible solution $(x^{(0)}, y^{(0)}, z^{(0)})$ of the primal and dual problems, and set the iteration number $S$=0. It should be noted that in this study the optimization variable $x$ in the primal problem denotes the power of the time-shiftable load $P_t^{CNLOAD}$; $y$ and $z$ are the optimization variables in the dual problem.

Step 2: Calculate the dual gap based on the current solution $(x^{(S)}, y^{(S)}, z^{(S)})$.

Step 3: If the dual gap is less than a pre-defined threshold $\varepsilon$, replace the optimal or near-optimal solution with the current solution and terminate the iteration process; otherwise, proceed to the next step (Liu et al., 2013). Note that, through large amounts of numerical simulations, the threshold is chosen as 10-5 in this study, since it gives satisfactory performances in most cases.

Step 4: Determine the movement direction $(\Delta x^{(S+1)}, \Delta y^{(S+1)}, \Delta z^{(S+1)})$, and update the current solution.

Step 5: Set $S$=$S$+1, and then go to step 2.

### 4.5 Determination of the final scheme

For purpose of determination of the final scheduling scheme from all the solutions obtained by solving the presented bi-level model, the following objective function of the joint optimization $F^{JO}$ is given by (Li et al., 2018b):

$$F^{JO} = \min \sqrt{\left(F_1^{JO} - F_1^{IO}\right)^2 + \left(F_2^{JO} - F_2^{IO}\right)^2} \tag{40}$$

where the superscript $IO$ denotes the independent operation without consideration of demand response, the superscript $JO$ represents the joint operation with consideration of demand response; $F_1^{JO}$ is the MG net operating cost considering both MG and user interests; $F_1^{IO}$ is the MG net operating cost considering only MG interests; $F_2^{JO}$ is the user cost considering both MG and user interests; $F_2^{IO}$ is the user cost considering only user interests.

Eq. (40) is a joint optimization formula that is employed to filter out the final scheduling scheme. Through alternate iterations of the upper- and lower- levels, multiple scheduling schemes and the $F_1^{JO}$, $F_1^{IO}$, $F_2^{JO}$ and $F_2^{IO}$ of each scheme can be obtained. According to Eq. (40), the scheme with the smallest $F^{JO}$ is considered as the final scheme.

### 4.6 Solution process

As shown in Fig. 2, the developed hybrid analytical-heuristic solution method consists of the following steps:

Step 1: Construct the upper-level model.
Step 2: Deterministic transformations of the chance constraint.
Step 3: Gain the MG scheduling model in the form of MILP.
Step 4: Preset the time-of-use electricity price.
Step 5: Calculate electricity prices using the proposed mechanism of real-time pricing.
Step 6: Use Jaya to solve the upper-level model.
Step 7: Check whether a solution is found. If found, proceed to the next step; otherwise, update load and confidence levels, and then go to step 3.
Step 8: Obtain the MG dispatch schemes in the current iteration.
Step 9: Build the lower-level model.
Step 10: Solve the lower layer via the IPM.
Step 11: Obtain the user electricity plan.
Step 12: Calculate $F_1^{JO}$ and $F_2^{JO}$.
Step 13: Check whether the iterations should be terminated.
Step 14: Calculate $F_1^{IO}$ and $F_2^{IO}$.
Step 15: Determine the final scheduling scheme.
Step 16: Output the final solution.

## 5. Case study

In order to evaluate the performances of the presented approach, it is tested on an improved MG test system originally designed by the Oak Ridge National Laboratory (ORNL) Distributed Energy Communications and Control (DECC) lab (Liu et al., 2017b), which is shown in Fig. 3. All numerical simulations have been run on a desktop personal computer (PC) with 8 GB RAM and two 2.4 GHz CPUs.

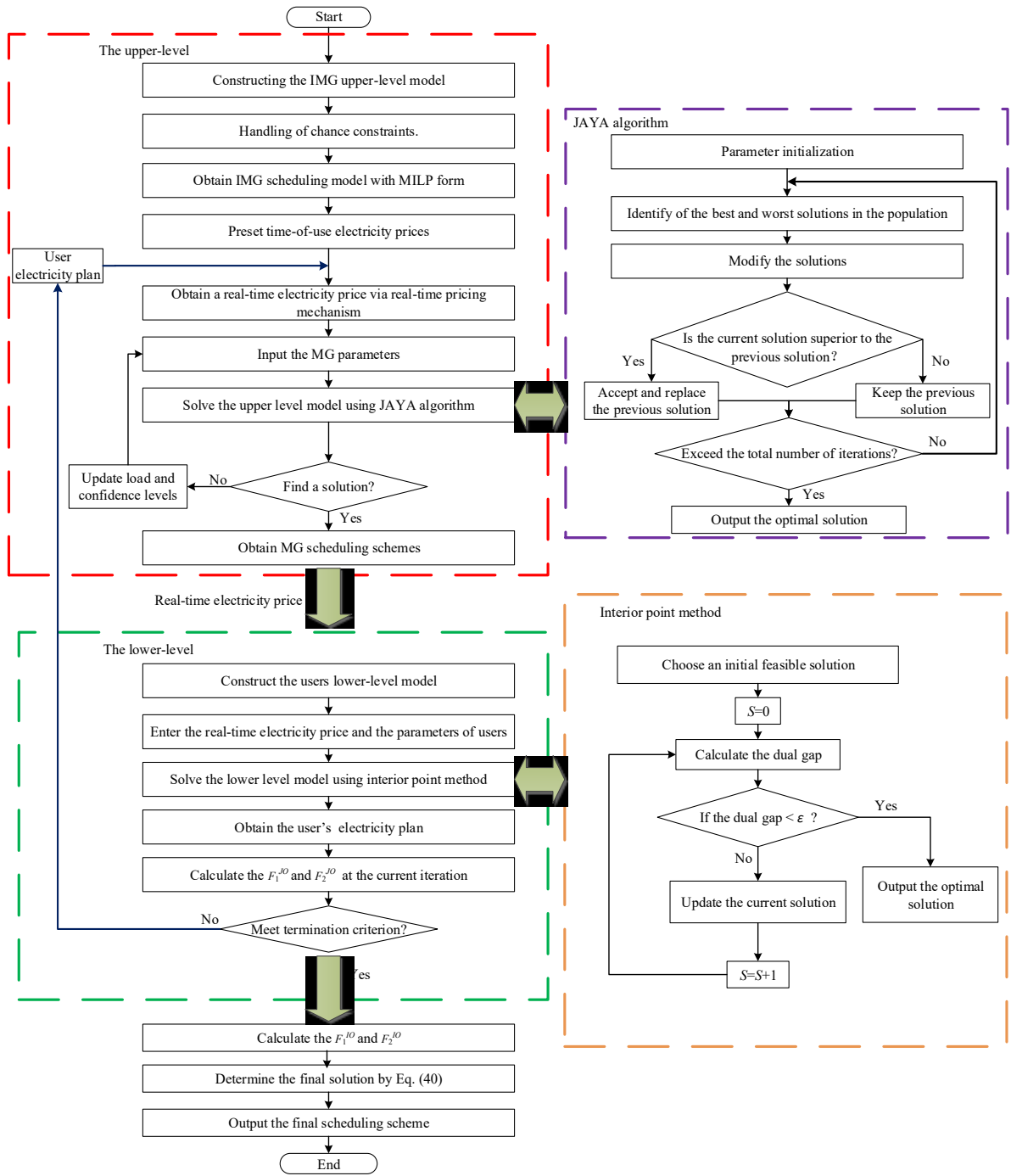

**Fig. 2.** Flow chart of the presented approach

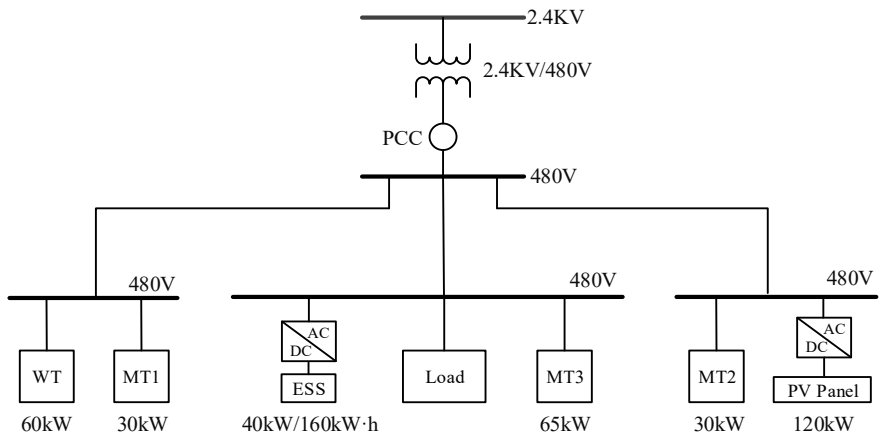

**Fig. 3.** Microgrid test system

## 5.1 Test system

As demonstrated in Fig. 3, the MG system mainly consists of 3 MT units, 1 WT unit, 1 PV panel, 1 ESS and load. Note that, the load here includes the time-shiftable load and the non-shiftable load.

The MT parameters are shown in Table 4.

**Table 4** MT Parameters

| MT number | $\zeta$($) | $\kappa_n$ ($) | $\Psi$($/kW) | $\varsigma$($/kW) | $P_{min}^{MT}$ (kW) | $P_{max}^{MT}$ (kW) |
|---|---|---|---|---|---|---|
| MT1 | 1.2 | 1.6 | 0.35 | 0.04 | 5 | 35 |
| MT2 | 1.2 | 1.6 | 0.35 | 0.04 | 5 | 30 |
| MT3 | 1.0 | 3.5 | 0.26 | 0.04 | 10 | 65 |

In this study, $\eta^{PV}$, $A^{PV}$ and $P_{max}^{PV}$ are 0.093, 1300 m², and 120 kW, respectively; $v_{in}$, $v_{Rated}$ and $v_{out}$ are 3 m/s, 15 m/s, and 25 m/s and the WT rated power output $P_{Rated}^{WT}$ is 60 kW, the battery parameters are as follows: $P_{max}^{DC} = P_{max}^{CH}$ =40 kW, $\eta^{DC} = \eta^{CH}$ =0.95, the maximum and minimum capacity of ESS are respectively 160 kW·h and 32 kW·h; the cost of ESS to provide reserve is 0.02 $/kW·h and the ESS charge/discharge prices are 0.3 $/kW·h and 0.5 $/kW·h, respectively (Li et al., 2019b); $P_{max}^{L}$ and $P_{REF}^{EL}$ are respectively set to 195 kW and 51.5 kW; $\omega_{REF}$ = 0.6 $/kW·h and $N_{iter,max}$ is 20. Regarding the parameters of the Jaya, the population size and the total number of iterations are respectively 100 and 1500.

## 5.2 Simulation scenario

As illustrated in Fig. 4, the simulation scenarios, including the load power, RG outputs and EL power, are utilized for the subsequent analysis.

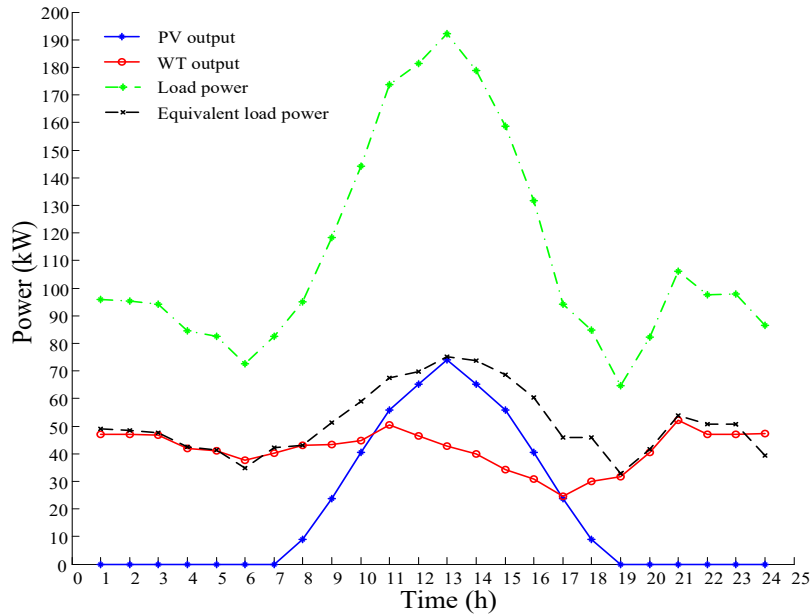

**Fig. 4.** The simulation scenarios

Fig. 4 shows that the joint power output of the two types of RGs can be obtained by using the addition-type-convolution, and thereby the expected value of the EL power is calculated via the subtraction-type-convolution operation between the joint power output and the load power.

## 5.3 Economic cost analysis

To properly evaluate the economic advantages of the proposed approach, the following strategies are adopted:

**Strategy 1:** optimal scheduling considering only MG interests. Specifically, strategy 1 only optimizes the upper-level model. This strategy guides users in the lower-level model to make electricity plans under the time-of-use electricity price, thereby ensuring that the net operating cost of the MG in the upper-level model is minimized;

**Strategy 2:** Coordinated scheduling of MG and users under demand response strategy. Specifically, strategy 2 optimizes the bi-level model. To balance the interests between MG and users, this strategy uses the designed real-time pricing mechanism to obtain a scheduling scheme through iterative alternations between the upper- and lower- levels, where electricity prices and electricity plans of the users act as the link of the two levels;

**Strategy 3:** optimal scheduling considering only user interests. To be specific, strategy 3 only optimizes the lower-level model. This strategy allows users to use electricity as much as possible during the lowest electricity price periods under the time-of-use electricity price without exceeding the amount of electricity provided by the MG, thereby minimizing the user cost in the lower-level model.

The above strategies are simulated with the load fluctuation $\partial$ =10%, the ratio of time-shiftable loads $\vartheta$ =0.2, the confidence level $\gamma$ =95%, $q$=2.5 kW. The costs of the IMG and users in the three optimization strategies are illustrated in Fig. 5.

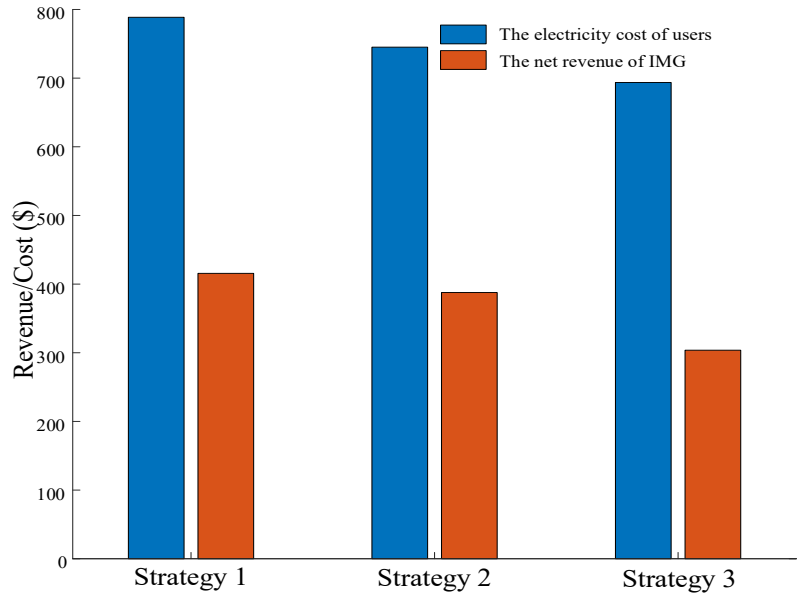

**Fig. 5.** Costs of the IMG and users in the three strategies

Fig. 5 illustrates that the results obtained from the three strategies are quite different from each other. Specifically, the scheme obtained in Strategy 1 gives the highest net revenues of the IMG and the highest user electricity cost; in Strategy 3, the obtained scheme generates the lowest user cost while the MG net revenue is the lowest. Note that the MG net revenue denotes the difference between the revenue and the MG operational cost. Compared with Strategies 1 and 3, the solution in Strategy 2 is capable of comprehensively taking into account the interests of both MG and user, which reduces the user electricity cost and increases the MG net revenues. Consequently, the above results demonstrate that the presented approach manages to balance the interests of multi-stakeholders by implementing demand response strategies.

Fig. 6 illustrates the initial real-time electricity price, the real-time electricity prices corresponding to the selected final scheme, time-of-use electricity price and the equivalent load in a scheduling cycle.

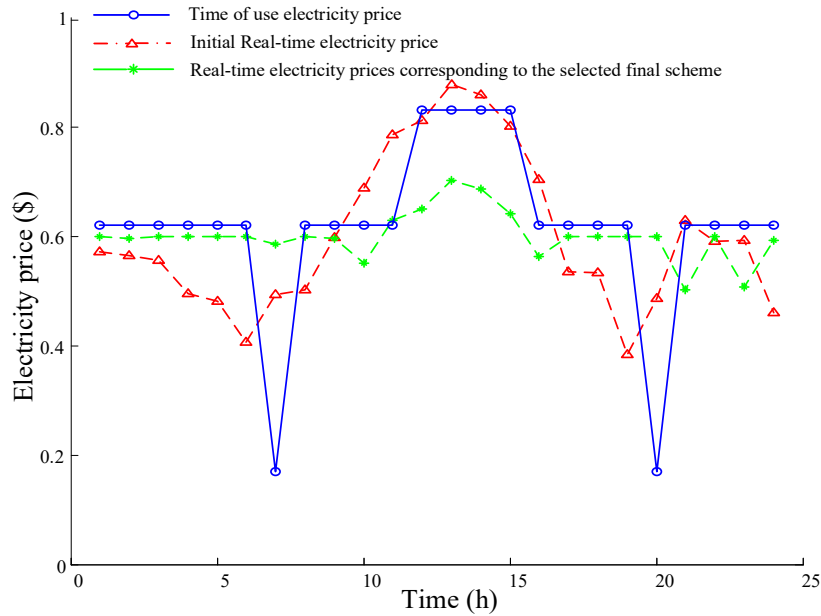

**Fig. 6.** Electricity prices

From Fig. 6, it can be observed that real-time electricity prices are superior to the time-of-use electricity price, achieving the effects of peak load shaving. Furthermore, the real-time price curve corresponding to the selected final scheme is significantly smoother than the curve of the initial real-time electricity price. Consequently, the effectiveness of the designed real-time pricing mechanism is verified.

### 5.4 Impacts of renewable uncertainties
#### 5.4.1 Relationship between MG operating cost and confidence level

To reasonably evaluate the changes in IMG operating costs as confidence levels rise, a set of simulation experiments under 10 confidence levels (50%, 55%, ..., 95%) are carried out, with the results shown in Fig. 7 below.

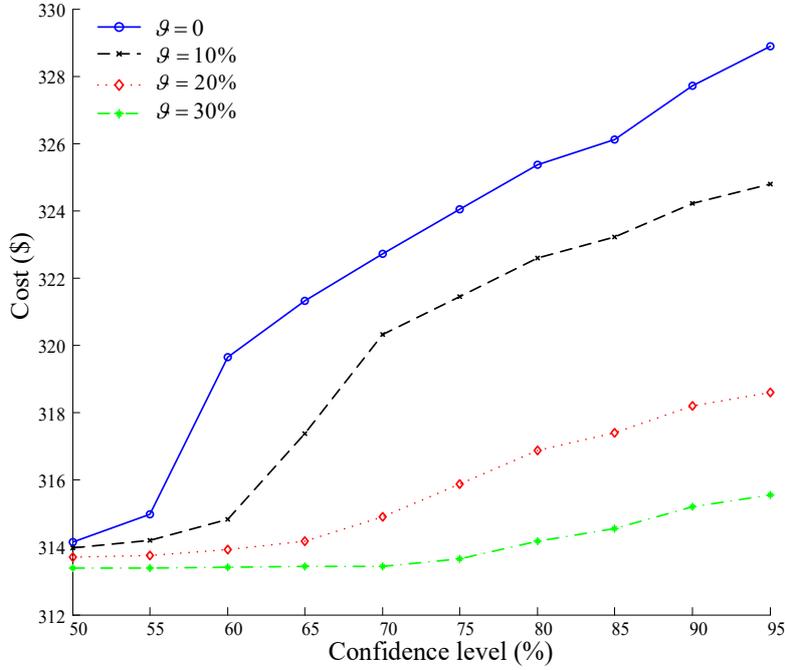

**Fig. 7.** MG operating cost as a function of confidence level

From Fig. 7, it can be illustrated that the operating costs of the IMG are closely related to confidence level $\gamma$. Specifically speaking, for a given time-shiftable load ratio $\vartheta$, the MG operating cost grows as the confidence level increases; while when the confidence level is fixed, the MG operating cost gradually decreases as the time-shiftable load ratio increases. Therefore, one can see that the proper selection of confidence level $\gamma$ is of great importance in balancing economy and reliability.

### 5.4.2 Relationship between spinning reserve capacities and confidence level

The proper configuration of spinning reserves is one of the most important means to ensure an active balance of a power grid. Increasing the spinning reserve will lead to an increase in IMG operating costs, which will have a serious impact on the economics of IMG operations. For this reason, given load fluctuation $\partial =10\%$ and the ratio of time-shiftable load $\vartheta=0.2$, the spinning reserve capacities are illustrated in Fig. 8.

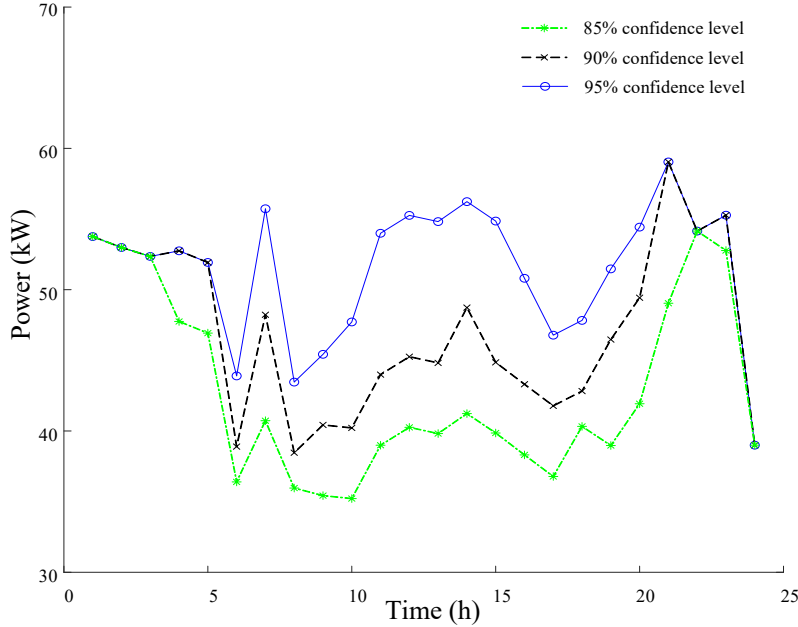

**Fig. 8.** Reserve capacities as a function of confidence level

As can be seen from Fig. 8, the proper selection of the confidence level is essentially a comprehensive decision-making issue regarding the reliability and economy of system operation. Specially speaking, with the confidence level increases, the reliability will correspondingly increase, but the economy will inevitably decrease; on the contrary, the reliability will decrease and the economy will increase. Therefore, it is critical to choose a reasonable confidence level to balance the economy and reliability. Furthermore, the CCP is thereby verified as an effective means to handle the uncertainties of both multiple renewable generations and loads.

### 5.5 Impacts of demand response

#### 5.5.1 Impact on the operations of microgrid

In order to examine the effectiveness of the DR strategy, this work analyzes two cases as follows:

**Case 1:** The electricity price exchanged between IMG and user is the time of use electricity price without DR strategy.
**Case 2:** The electricity price exchanged between IMG and user is the real-time price considering demand response. Herein, renewable uncertainty and demand response are coordinated under the mechanism of real-time pricing.

Based on previous research, the time-of-use electricity prices are set as follows: during peak period (11:00-15:00), the price is 0.83 $/kWh; Off-peak period (06:00-07:00,18:00-19:00) the price is 0.17 $/kWh; it is taken as 0.62 $/kWh in other periods.

Given the load fluctuation $\partial$ =10%, the ratio of time-shiftable load $\vartheta$ =0.2 and the confidence level $\gamma$ =95%, comparative tests have been carried out to compare the obtained scheduling schemes with and without consideration of demand response, and the test results are respectively shown in Figs. 9 and 10.

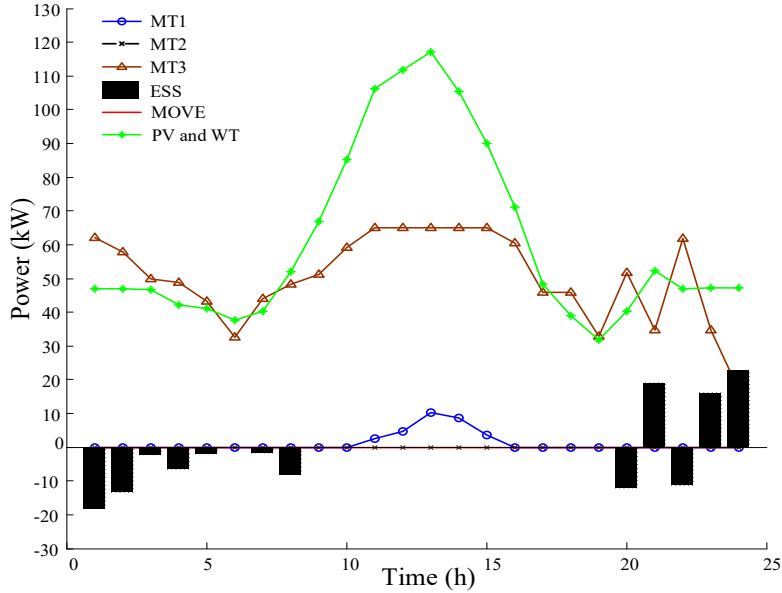

**Fig. 9.** Scheduling strategy without demand response

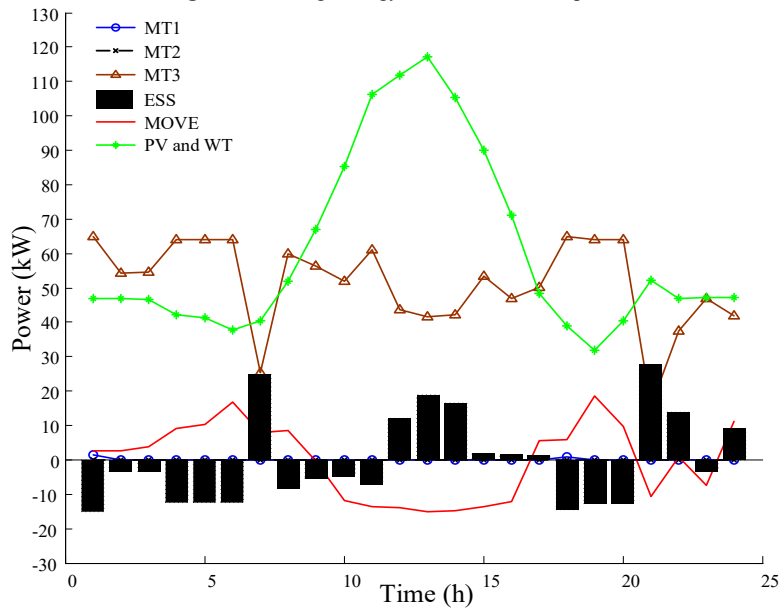

**Fig. 10.** Scheduling strategy with demand response

From Figs. 9 and 10, the following important facts can be found. (1) DR enables consumers to play an important role in MG operations by actively adjusting their electricity plans. More specifically, time-shiftable loads can shift out to a large extent during the peak load period and shift during the off-peak period by introducing DR, as shown in Fig. 10. (2) Regarding the MTs, the MT outputs and the corresponding fuel cost with consideration of DR can be significantly reduced during the peak load period. (3) As far as the ESS is concerned, the ESS charge-discharge frequency in Fig. 10 is significantly higher than that in Fig. 9. This is because the economic operation of MG is related to the battery cost of charging and discharging. Therefore, it can be concluded that peak load shaving can be achieved through the coordination between the renewable uncertainties and demand response, balancing the interests of both MG and users.

### 5.5.2 Impact on load powers

The equivalent load powers with and without considering demand response are shown in Fig.11.

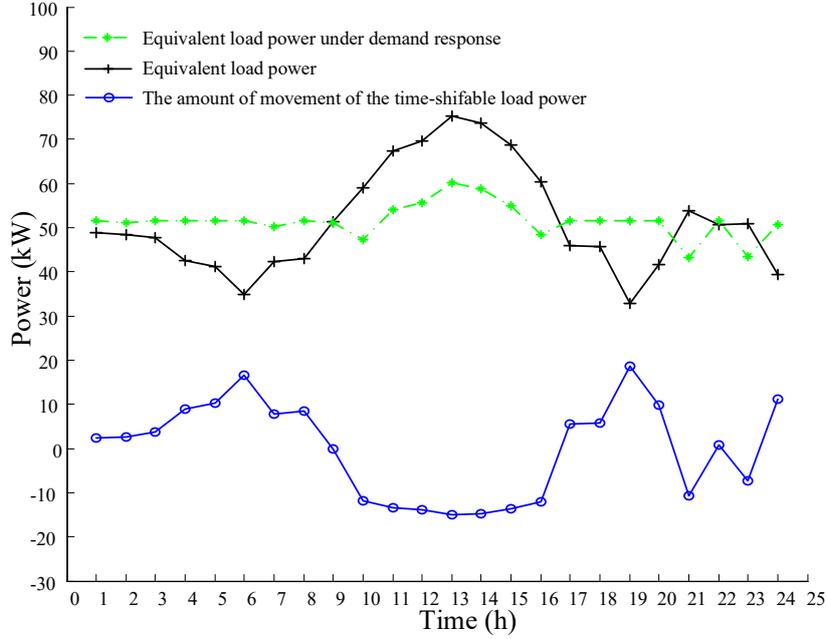

**Fig. 11.** Load powers with and without considering demand response

Fig. 11 suggests that the equivalent load power with considering demand response is obviously superior to that without considering demand response, which reflects in its smoother trend. The reason for this is that due to demand response at peak load periods the time-shiftable load shift out some loads, while at the off-peak periods the time-shiftable load shift in a part of loads. In this way, users can be actively taken part in the economic operation of MG, and peak load shaving can be realized.

### 5.6 Comparison with other solution methods

In order to reasonably evaluate the proposed method, two algorithms are used for comparative analysis. Firstly, we use the CPLEX solver to solve the upper- and lower-level models separately. Secondly, the commonly-used hybrid intelligent algorithm (HIA), which combines particle swarm optimization (PSO) algorithm with Monte Carlo simulations (Li et al., 2018b), has been adopted to address the model, and a comparative test has also been carried out. Regarding the HIA, the population size of the PSO is 20, and the maximum number of iterations is 200; the number of random variables in the Monte Carlo simulations is set to 500 (Li et al., in press-b). Note that the results are the average values of 20 independent runs due to the randomness of the HIA (Li et al., 2019b).

A comparison of the three solution methods has been carried out with the results illustrated in Table 5.

**Table 5** Comparison results of the proposed method and other alternatives

| Solution method | $\vartheta$ (%) | Economy ($) | | Calculation time (s) |
|---|---|---|---|---|
| | | MG net revenue | User cost | |
| HIA | 0 | 392.1 | 816.3 | 327.9 |
| | 20 | 349.7 | 793.2 | 352.3 |
| CPLEX | 0 | 389.7 | 824.8 | 71.1 |
| | 20 | 346.6 | 807.6 | 84.8 |
| **Proposed method** | 0 | **409.4** | **780.8** | **33.9** |
| | 20 | **387.8** | **745.1** | **37.4** |

From Table 5, it can be seen that the proposed method is significantly better than the HIA and CPLEX in terms of economy and calculation time. As far as the economy is concerned, given $\vartheta$ =20%, the MG net revenue obtained by the proposed method is 10.9% and 11.9% higher than that of HIA and CPLEX, while the user cost of our method is 6.1% and 7.7% lower than that of HIA and CPLEX; regarding the calculation time, the time of our method is reduced by about 90% and 60% compared with the HIA and CPLEX, respectively. Therefore, it can be safely concluded that the proposed Jaya-IPM method is suitable for the bi-level programming problem in this study.

### 6. Conclusion

With the intensification of the energy crisis and people's increasing concerns about sustainable energy supply, microgrids are becoming an indispensable means for addressing such energy and environmental issues. From the perspective of multi-stakeholders, this work proposes a stochastic scheduling model for demand response-enabled microgrids with renewable generations and develops a hybrid analytic-heuristic solution approach. Numerical simulation results have verified that the following main conclusions:

(1) The proposed scheduling approach is capable of coordinating demand response and renewable generations for microgrid scheduling, achieving peak load shaving. In particular, during the peak load period, the reduced power demand reduces the spinning reserves that are used for dealing with the uncertainties of renewable generations; while during the off-peak period, the increased power demand can consume more renewable power outputs.

(2) The developed analytic-heuristic hybrid algorithm Jaya-IPM is able to solve the proposed scheduling model accurately and quickly. With regard to the economy, the MG net revenue of the proposed method is 10.9% and 11.9% higher than that of HIA and CPLEX, while the user cost is 6.1% and 7.7% lower than that of the HIA and CPLEX; in terms of the calculation time, the calculation time of the proposal is decreased by about 90% and 60% compared with the HIA and CPLEX.

(3) The test results demonstrate that, through the coordination of demand response and uncertain renewable generations, this approach not only guides users to actively take part in microgrid scheduling but also significantly mitigates the effects of renewable generation uncertainties, which provides a way of promoting the ongoing low-carbon transition towards sustainable production.

In addition, there are some limitations in this work, which are manifested in the following two aspects: on the one hand, the losses of ESS capacity and the transmission loss of power supply are not considered in this paper; on the other hand, there is no further research on the participation and satisfaction of different types of users with DR strategies.

In future research, game theory-based demand response can be introduced to develop a more realistic microgrid scheduling model. Besides, as data privacy issues are increasingly concerned (Li et al., in press-c), it is interesting to study to achieve a privacy-preserving energy scheduling of microgrids. Although the size of this studied case is a park-level microgrid, our method can deal with a larger case due to its good scalability. Another interesting topic is to extend the presented method to multi-objective energy management of a microgrid with uncertain renewable generations by combining multi-objective optimization with integrated decision making (Li et al., 2018e).

## Acknowledgements

This work is partly supported by the Natural Science Foundation of Jilin Province, China under Grant No. YDZJ202101ZYTS149.